\documentclass[pdftex,10pt,a4paper,twoside,twocolumn,fleqn]{article}
\usepackage{graphicx}
\usepackage{color}

\usepackage{csquotes}
\MakeOuterQuote{"}

\patchcmd{\maketitle}{\@fnsymbol}{\@alph}{}{}

\usepackage[pdftex,
    bookmarks,
    bookmarksopen=true,
    bookmarksnumbered=true,
    pdfauthor={D.~P.~Mohr, C.~A.~Knapek, P.~Huber, E.~Zaehringer},
    pdftitle={Particle Detection Algorithms for Complex Plasmas},
    pdfkeywords={},
    colorlinks, linkcolor=blue, urlcolor=blue
]{hyperref}

\usepackage{multirow}

\usepackage{mathpazo}
\usepackage[T1]{fontenc}
\usepackage{textcomp}
\usepackage{amssymb}
\usepackage{amsmath}
\usepackage{natbib}
\usepackage[all]{xy}

\definecolor{violet}{rgb}{1,0,1}
\definecolor{french_rose}{RGB}{246,74,138}
\definecolor{peach}{RGB}{255,229,180}
\definecolor{royalblue}{RGB}{65,105,225}
\definecolor{mint_green}{RGB}{150,250,150}

\newcommand{\frenchrose}{pink ({\color{french_rose}$\blacksquare$}) }
\newcommand{\peach}{yellow ({\color{peach}$\blacksquare$}) }
\newcommand{\royalblue}{blue ({\color{royalblue}$\blacksquare$}) }
\newcommand{\mintgreen}{green ({\color{mint_green}$\blacksquare$}) }

\newcommand{\R}{\ensuremath{\mathbb{R}}}
\newcommand{\abs}[1]{\left|#1\right|}

\newcommand{\erf}[1]{\operatorname{erf}{#1}\,}
\newcommand{\SNR}{\operatorname{SNR}\,}
\newcommand{\ceil}[1]{\left\lceil{#1}\right\rceil}
\newcommand{\floor}[1]{\left\lfloor{#1}\right\rfloor}
\newcommand{\Z}{\ensuremath{\mathbb{Z}}}
\newcommand{\figref}[1]{\autoref{#1}}
\newcommand{\secref}[1]{\autoref{#1}}

\newcommand*{\doi}[1]{\href{http://dx.doi.org/#1}{doi: #1}}

\begin{document}
\title{Particle Detection Algorithms for Complex Plasmas}
\author{Daniel~P.~Mohr \thanks{Deutsches Zentrum f{\"u}r Luft- und Raumfahrt e.~V., Institut f{\"u}r Materialphysik im Weltraum, 82234 Wessling, Germany. \href{https://complex-plasmas.dlr.de/}{complex-plasmas.dlr.de}} \and
Christina~A.~Knapek\footnotemark[1] \and
Peter~Huber\footnotemark[1] \and
Erich~Zaehringer\footnotemark[1]}
\date{\today}

\maketitle

\section*{Abstract}
In complex plasmas, the behavior of freely floating micrometer sized particles is studied. The particles can be directly visualized and recorded by digital video cameras. To analyze the dynamics of single particles, reliable algorithms are required to accurately determine their positions to sub-pixel accuracy from the recorded images. Typically, straightforward algorithms are used for this task.
Here, we combine the algorithms with common techniques for image processing. We study several algorithms and pre- and post-processing methods, and we investigate the impact of the choice of threshold parameters, including an automatic threshold detection.
The results quantitatively show that each algorithm and method has its own advantage, often depending on the problem at hand. This knowledge is applicable not only to complex plasmas, but useful for any kind of comparable image-based particle tracking, e.g.\ in the field of colloids or granular matter.


\section*{Introduction}
Particle detection in digital images is a crucial first step in the analysis of many-particle systems in the case that individual particles can be detected by direct optical measurements. Efforts to optimize particle detection can be found in a wide range of fields: in biophysics, single particle tracking is used to study the motion of 
particles (e.g.\ proteins, molecules or viruses) involved in cell membrane and intracellular activities \citep{Saxton:1997,Sbalzarini:2005,Chenouard:2014}. 
Particle detection and tracking from optical measurements is utilized in granular matter research \citep{Tsai:2004,Harrington:2014}, and in colloidal physics, where the dynamics of systems of nano- to micrometer sized particles can be investigated by analyzing single particle motion from direct video microscopy \citep{Crocker:1996,Leocmach:2013}.

Complex plasmas \citep{Fortov:2005,Morfill:2009,Ivlev:2012} consist of micrometer sized particles injected into a low-temperature plasma composed of electrons, ions and neutral gas. These particles are large enough to be visible to digital cameras with an appropriate optics, and provide an excellent opportunity to study fundamental dynamics on the kinetic level of individual particles.
In contrast to colloids, where particles are embedded in a viscous medium and therefore over-damped, complex plasmas are virtually undamped, and the time scale of dynamical processes is short and therefore easily accessible. The particles are typically illuminated with a sheet of laser light, and the reflected light can be observed with digital cameras. Since the particle distances are large (with a magnitude of several hundreds of micrometers), individual particles can be observed directly as mostly disjunct small groups of illuminated pixels on the camera sensor. 

From those ``blobs'' of pixels, particle positions can be determined to sub-pixel accuracy -- a necessity for the study of dynamics of single particles -- with an adequately chosen algorithm.

By detecting individual particles, and tracing them through consecutive images (this is possible if the particle displacement between two images is small enough to allow for a unique assignment), velocities can be obtained. This method is called Particle Tracking Velocimetry (PTV), and has the advantage of more precise velocity measurements \citep{Pereira2006} in contrast to Particle Image Velocimetry (PIV)\citep{Williams2016}, where only spatially averaged velocity vectors are obtained, especially in particle clouds too dense for single particle detection.

Complex plasmas are three-dimensional systems, and currently the interest in 3D optical particle diagnostics is growing \citep{Jambor:2016,Melzer:2016}. To triangulate the real position of a particle in 3D space, additional requirements are imposed on particle detection algorithms. Hence, we are also looking for algorithms for the detection of particles which are nearby each other on the image plane due to their overlapping motion in different layers. These algorithms can also be useful for particle tracking in systems with a high packing density.

With the methods presented in this paper we show that the commonly acquired accuracy can be exceeded without unnecessary increasing the complexity of the procedure. This is accomplished by combining simple image pre- and post-processing procedures with an improved version of the commonly used algorithm for blob detection, and to some extent by applying automatic threshold detection.

Usually, straightforward algorithms are used for blob detection \citep{Crocker:1996, Feng:2007, Ivanov:2007}, which is justified by the simple search feature and the typically low image noise. We show that this algorithm can be improved by generalizing it to blobs being not necessarily simply connected sets of pixels. Other more complex blob detection algorithms, such as SimpleBlobDetector \citep[SimpleBlobDetector]{opencv_library} or MSER \citep{Matas:2002}, did not turn out to be satisfactory in our case.

Though some of the techniques are well-known, a combination of them as well as an investigation of their individual influence on the accuracy of particle detection was not performed elsewhere to such an extent, especially for the typical particle shapes obtained in complex plasma experiments (for example, \citet{Feng:2007} is mostly involved in examining one particular core algorithm without pre- and post-processing, while \citet{Ivanov:2007} investigated some methods for pre- and post-processing, but does not combine the methods in the result).

Here, we not only investigate pre-processing, particle detection and post-processing in combination, but also take into account particle sizes and several kinds of image noise in our results. Additionally, we introduce Otsu binarization as an automated procedure. We also show that the choice of methods strongly depends on the image features (e.g. noise).

The paper is organized as follows: After a description of the general approach in \secref{section:General Approach}, \secref{section: Simulated Images} shows how the artificial images are generated to test the quality of the algorithms. In sections \ref{section:Preprocessing (Image Processing)}, \ref{section:Moment Method (Blob Detection)} and \ref{section:Fitting Postprocessing} the different steps of image processing and particle detection are presented in more detail, followed by some examples in \secref{section:Examples}. Finally, the conclusion summarizes the paper in \secref{section:Conclusion}.

\section{General Approach}\label{section:General Approach}

The process of obtaining particle coordinates from experimental data (images) can be divided into the following necessary steps:

\begin{description}
\item [Image acquisition] Get the image from real world.
\item [Image processing] Prepare/enhance the image (e.g.\ by filtering).
\item [Blob detection] Identify particle positions.
\item [Postprocessing] Enhance found positions of particles (e.g.\ fitting).
\end{description}

Each step is an own field of research. In the following, they are explained in the depth necessary for this work.

\subsection{Image Acquisition}
Image acquisition is part of the experiment, and is only mentioned here for completeness, since the details of the experimental procedures go beyond the scope of this paper. To get good images, we need a proper illumination of the particles, a matched optical system, and an appropriate camera with an applicable storage system.

In this step, image noise is introduced. The sources are manifold, e.g.\ thermal behavior of the camera chip, noise of the involved electronics, defect pixels or radiation influencing the complete system. The noise causes uncertainties, which can be abstracted as additive white Gaussian noise and salt and pepper noise, superimposed on the pixels.

In the following, we assume a camera giving $8$ Bit gray scale images.

\subsection{Image Processing}
Preparing the image is a task extremely dependent on the blob detection algorithm to be chosen for the next step, e.g.\ an algorithm using edge detection will not work well if the edges of the blob are destroyed by applying a smoothing filter. In that case, a sharpening filter would be preferable.

One particle can be seen approximately as a point source of light, and the point spread function describes what we can expect to see on the image sensor. The point spread function defines how an ideal point source will be mapped by a system of optical components. In the case of point-like particles, the Airy disc \citep{Airy:1835} gives a good approximation of this mapping. 

Optical side lobes of the point spread function can be reduced by a Hanning amplitude filter (a convolution with the Hann function) \citep{Kumar:2013}. The Hann function, visualized in \figref{fig:example shapes}, with the parameter $N$ for a point $r$ is given by:
\begin{align*}
\hspace{-7mm}w(r) = \left\{ \begin{array}{ll}
    \frac{1}{2} - \frac{1}{2} \cos{\left( \frac{ 2 \pi \left(r-\frac{N-1}{2}\right)}{N-1} \right)} & \mbox{\hspace{-2mm}if } \abs{r} \leq \frac{N-1}{2} \\
    0 & \mbox{\hspace{-2mm}else.} \\ \end{array}\right.
\end{align*}

The parameter $N$ influences the width of the window. The Hanning amplitude filter is in principal a low-pass filter. This kind of filter passes signals with a spatial frequency less than the (user chosen) cutoff frequency, and can therefore reduce high-frequency image noise, e.g.\ Gaussian white noise.

This filter can easily be implemented by using template matching from the library opencv \citep{opencv_library}.

\begin{figure}
  \includegraphics[width=0.49\textwidth]{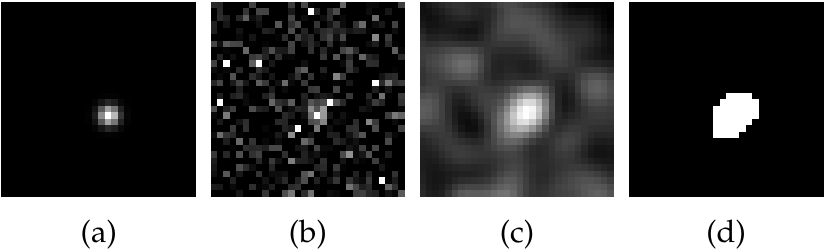}
  \caption{From left to right:
    (a) Image of a particle of size $\sigma_{x,y} =1$. In (b) high noise is added (Gaussian noise with $\SNR =5$, and salt and pepper noise with a probability of $0.5 \mbox{ \%}$), (c) the noisy image is then filtered by a Hanning amplitude filter with $N=5$, (d) and finally the filtered image is clustered by Otsu's method.
  }\label{fig:example of hann filtered image}
\end{figure}

In \figref{fig:example of hann filtered image}, an example shows the effect of a Hanning amplitude filter.

Of course it is in general a good idea to use combined low-pass and high-pass filtering. High-pass filtering does the opposite to a low-pass filter: pass spatial frequencies below a cutoff and thus reduce image noise such as large-scale intensity gradients. But a high-pass filter can mask the behavior of a low-pass filter -- e.g.\ the blurring of a low-pass filter would be reduced by a high-pass filter. Since we want to investigate the effect of specific filters, we do not want this masking. Usually, we do not observe low frequency noise in our images, and therefore omit high-pass filtering in this paper.

In general it should be mentioned, that \citet{Crocker:1996} and \citet{Ivanov:2007} use a simple but effective filter, which behaves similar to a high-pass filter. They subtract a background calculated by a convolution with a boxcar kernel (moving average) from the image after low-pass filtering by a convolution with a Gaussian kernel.

\subsection{Blob Detection}\label{section:Blob Detection}
In the complex plasma community, a typical approach for blob detection is the moment method \citep{Feng:2007, Ivanov:2007}, which is a simplified version of the approach by \citet{Crocker:1996} \footnote{You can find many implementations in the internet, e. g.: \href{http://www.physics.emory.edu/faculty/weeks/idl/}{www.physics.emory.edu/faculty/weeks/idl}}:

\begin{enumerate}
\item Find connected pixels brighter than a threshold (a particle).
\item Calculate center of every particle (position of a particle).
\end{enumerate}

In the literature \citep{Feng:2007, Ivanov:2007}, connected pixels are assumed to be a set, which is simply connected. More general, we now define a set of pixels $P_i$ belonging to a particle as:
\begin{alignat*}{2}
P_i \subset &\left\{ p: I(p)>I_{threshold} \right\}\\
\end{alignat*}
\begin{alignat*}{2}
  \mbox{with: }& \left(\left|P_i\right| = 1\right) \lor \\
  & \quad \left( \forall  p_j \in P_i: \right.\\
  & \qquad \left. \exists p_k \in P_i \setminus \left\{p_j\right\}:d(p_j, p_k)<r \right)\\
  & \forall  p_j \in P_i: \\
  & \quad \forall p_k \in \left\{ p: I(p)>I_{threshold} \right\} \setminus P_i:\\
  & \qquad d(p_j, p_k) \geq r\\
  &\left|P_i\right| \geq m_p\\
& x_d\left(P_i\right) \geq m_x \\
& y_d\left(P_i\right) \geq m_y \\
&\frac{\left|P_i\right|}{ x_d\left(P_i\right) \times y_d\left(P_i\right) } \geq m_d\\
& \frac{1}{\left|P_i\right|} \sum_{p \in P_i}^{}{I(p)} \geq m_{bd}\\
\end{alignat*}
\begin{alignat*}{2}
\mbox{with: }& x_d(P_i) := \max\left\{x\left(p_j\right)-x\left(p_k\right): \right.\\
& \qquad \qquad \qquad \qquad \left. p_j, p_k \in P_i\right\}\\
& y_d(P_i) := \max\left\{y\left(p_j\right)-y\left(p_k\right): \right.\\
& \qquad \qquad \qquad \qquad \left. p_j, p_k \in P_i\right\}\\
\end{alignat*}

Here, $P_i$ is a set of pixels and represents the particle with the number $i$, $I(p)$ is the intensity of the pixel $p$, $I_{threshold}$ is the intensity of the threshold, $\left|P_i\right|$ denotes the cardinal number of $P_i$, $d(p_j,p_k)$ is the distance of the two pixels $p_j$ and $p_k$, $r$ is a search radius, $m_p$ is the minimum number of pixels a particle needs to be composed of, $x(p)$ is the $x$ coordinate of $p$, $y(p)$ is the $y$ coordinate of $p$, $m_x$ is the minimum length in $x$ direction in pixel, $m_y$ is the minimum length in $y$ direction in pixel, $m_d$ is the minimum density of a particle (density being the total number of pixels weighted by the area of the smallest rectangle envelope of $P_i$), $m_{bd}$ is the minimum brightness density. The brightness density is defined as the sum of all intensity values of the pixels in $P_i$, weighted by the total number of pixels in the set $P_i$.

The parameter $r$ allows to identify sets of pixels as a particle $P_{i}$ even if those pixels are not directly connected. For example, setting $r = 1$ leads to a simply connected set as used in the mentioned literature, while setting $r = 1.5 > \sqrt{1^2+1^2}$ (assuming quadratic pixels with side length $1$) allows pixels in $P_{i}$ to be connected only by a corner. 
For larger values of $r$, the pixels in the set $P_{i}$ do not need to be simply connected at all. This can be used for compensation of pepper noise or intensity jitter.
In addition, to be recognized as separate particles, the shortest distance between the particle contours of two neighboring particles must be $\ge r$.

The center $\left(x_c(P_i), y_c(P_i)\right)$ can be calculated using the pixel positions and --- as often done in the mentioned literature --- the brightness of them:
\begin{alignat*}{2}
x_c(P_i) = \frac{\sum_{p \in P_i}^{}{x(p) \left(I(p) - I_{base}\right)}}{\sum_{p \in P_i}^{}{\left(I(p) - I_{base}\right)}}\\
y_c(P_i) = \frac{\sum_{p \in P_i}^{}{y(p) \left(I(p) - I_{base}\right)}}{\sum_{p \in P_i}^{}{\left(I(p) - I_{base}\right)}}
\end{alignat*}
Here, $I_{base}$ gives an offset. In \citet{Feng:2007} this offset is discussed and it is recommended to use $I_{base} = I_{threshold}$ to reduce the error.

Other blob detection algorithms (\citet[SimpleBlobDetector and MSER]{opencv_library}) were tested, but proved to be unreliable and could only detect some of our largest particles. Since those algorithms increase the complexity and computation time without reaching the quality of our proposed blob detection method for the small particle images prevalent in complex plasmas, they were not investigated further.

\subsection{Postprocessing}
\begin{figure}
  \includegraphics[width=0.49\textwidth]{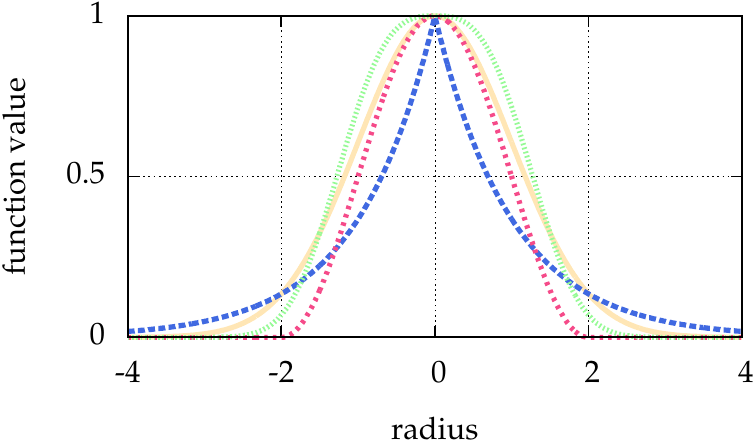}
  \caption{Shown are: the Hann function with $N=5$ in \frenchrose, and the generalized Gaussian point spread function with $p=1$ in \royalblue,  $p=3$  in \mintgreen and  $p=2$ in \peach. The latter is identical to the normal distribution. The width is $\sigma=1$ for all cases. With different parameters $p$, the generalized Gaussian is able to mimic different particle shapes, which can e.g.\ result from defocused images.}\label{fig:example shapes}
\end{figure}
Since the blob detection is not an exact deconvolution, we are bound to have errors. To overcome this, we can fit a function to the approximate particle coordinates obtained from the blob detection. We now use the concept of a particle as an approximate point source of light, and the subsequent description of $P_i$ as a point spread function similar to the Airy disc \citep{Airy:1835}. The latter can be approximated by a Gaussian or a generalized Gaussian point spread function \citep{Claxton:2008} (see \eqref{eq:generalized Gaussian point spread function} in \secref{section:Fitting Postprocessing}), visualized in \figref{fig:example shapes}.

In our procedure, we choose a generalized Gaussian point spread function and fit it to the approximate coordinates from the blob detection.

\section{Simulated Images}\label{section: Simulated Images}
To test our implementation we need well-defined, artificial images of particles. Here, the use of artificial images with well-defined particle positions is crucial to be able to calculate the deviation of tracked position to real position and thus to quantify the quality of our algorithms. The images are modelled after real-world experimental images of complex plasmas recorded by optical cameras.

The particles are represented by a bivariate normal distribution with a correlation of $0$:
\begin{align*}
\frac{1}{2 \pi \sigma_x \sigma_y}
\exp{\left( \frac{-1}{2}
 \left(\frac{(x-\mu_x)^2}{\sigma_x^2}+
 \frac{(y-\mu_y)^2}{\sigma_y^2}\right)
\right)}
\end{align*}

Since $(2 \pi \sigma_x \sigma_y)^{-1}$ is just a constant factor, it can be ignored. Furthermore, the image is scaled to values between $0$ and $1$.

In a real-life camera image, the brightness of a pixel is the integration over time and space. Therefore, we integrate the intensity values over one pixel:
\begin{align*}
&\int_{x_1}^{x_2}{
 \int_{y_1}^{y_2}{}}\\
& \exp{\left( \frac{-1}{2}
  \left(\frac{(x-\mu_x)^2}{\sigma_x^2}+
   \frac{(y-\mu_y)^2}{\sigma_y^2}\right)
  \right)}
 {dy}
{dx}
\end{align*}
\begin{align*}
        = &\left(\int_{x_1}^{x_2}{
        \exp{\left(\frac{-(x-\mu_x)^2}{2\sigma_x^2}\right)}
        dx}\right)\\
        &\times\left(\int_{y_1}^{y_2}{
        \exp{\left(\frac{-(y-\mu_y)^2}{2\sigma_y^2}\right)}
        dy}\right)
\end{align*}
\begin{align*}
        = &\frac{\pi \sigma_x \sigma_y}{2}
        \left(\erf{\left(\frac{x_1-\mu_x}{\sqrt{2} \sigma_x}\right)}-
          \erf{\left(\frac{x_2-\mu_x}{\sqrt{2} \sigma_x}\right)}
        \right)\\
        &\times \left(\erf{\left(\frac{y_1-\mu_y}{\sqrt{2} \sigma_y}\right)}-
          \erf{\left(\frac{y_2-\mu_y}{\sqrt{2} \sigma_y}\right)}
        \right)
\end{align*}

Again, the constant factor $(\pi \sigma_x \sigma_y)/2$ can be ignored, because the image is rescaled in the end. This procedure is repeated for each pixel.

\begin{figure}
  \includegraphics[width=0.49\textwidth]{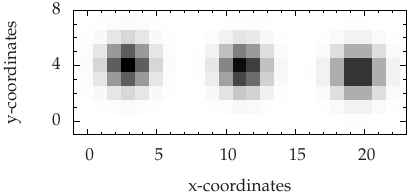}
  \caption{Simulated particles with $\sigma_x = 1 = \sigma_y$ without image noise. For better visibility the image is shown in inverted colors. The coordinates $(3, 4)$ of the left particle are centered in a pixel, the coordinates $(19.5, 3.5)$ of the right particle are exactly centered between two pixels, and the coordinates $(11.25, 3.8)$ of the middle one are chosen arbitrarily.}\label{fig:example particle sigma=1}
\end{figure}

Examples for artificial particle images are given in \figref{fig:example particle sigma=1}. The figure also illustrates the impact of the given sub-pixel location of the particle center on the intensity distribution.

To be able to describe the strength of the noise by one single parameter, we create an additive white Gaussian noise (AWGN) with a mean of $0$ and a standard deviation of $1$.
We can scale the noise to the image by a signal to noise ratio $\SNR$ with $B$ a matrix representing the noise free image, $B_{noise}$ a matrix representing the noise and $B_{noisy}$ a matrix representing the image with noise:
\begin{align*}
B_{noisy} = \max{\left\{0, \min{\left\{1,B + \frac{1}{\SNR} B_{noise}\right\}}\right\}}
\end{align*}

With this widely-used, simple noise model (e.g.\ it is often used in information theory \cite{Shannon:1948, Proakis:2001}), we can create a noise which behaves roughly similar to the thermal noise of camera sensors. In \citet[pages 43--44]{Pitas:2000}, this approach of cutting values is used to generate additive Laplacian noise\footnote{Because of a simple pseudo random number generator it was necessary to use a Laplacian instead of a Gaussian distribution in \citep[pages 43--44]{Pitas:2000}.}.

Our simple $\SNR$ is consistent with the well-known Rose criterion (\citet[page 97]{Rose:2013}), which states that a $\SNR$ of at least $5$ is necessary for a reliable detection. Due to this fact, \figref{fig:noise with snr} does not show bars for $\SNR = 5$.

By setting pixel intensities to $0$ or $1$ with a given probability, we can add salt and pepper noise. This kind of noise simulates defective pixels usually present on typical camera sensors. Pixels can appear dark (``pepper'') or bright (``salt''), regardless of the exposure, e.g.\ due to errors in the pixel electronics. 
Bright pixels are easy to detect by taking dark-frame images (e.g.\ an image taken with covered lens), dark pixels can be detected with more effort by taking gray images. If a list of defective pixels is available, some cameras are able to correct these listed pixels by averaging the surrounding ones.

Though the occurence of excessive salt and pepper noise in an experimental setup should normally lead to an exchange of hardware, there are situations in which this is not an option. Good examples are experimental instruments in remote locations not accessible to technicians, e.g. satellites or sealed experimental devices on a space station, such as complex plasma microgravity research facilities (PK-3 Plus \citep{Thomas:2008}, PK-4 \citep{Pustylnik:2016}). Here, cameras are exposed to higher levels of radiation, and pixel deterioration, causing salt and pepper noise, becomes an issue. To still obtain good scientific results over an extended period of time, one needs to handle such noise sources adequately during data analysis as long as it is feasible.

\section[Preprocessing (Image Processing)]{Preprocessing\\(Image Processing)}\label{section:Preprocessing (Image Processing)}
Image preprocessing is not restricted to the use of general filters preserving the brightness distribution of particles, but can be extended to procedures for e.g.\ threshold detection, especially with regard to the requirements of the moment methods.

In the first step, the moment method needs a separation of the pixels belonging to particles, and pixels composing the background. Since our images represent particles illuminated with a laser, we can assume to have a bi-modal histogram.

This can be clustered for example by Otsu's method \citep{Otsu:1979}. This method separates the histogram of the image in $2$ classes -- below and above the threshold -- with the goal to minimize the variances of both classes. This leads to a maximal interclass variance. The image is then binarized according to the classes --- pixels of one class are usually shown as white, and those belonging to the other class as black.

There are other thresholding techniques available (for an overview see e.g.\ \citet{Sezgin:2004}). We use Otsu's method since it is in the top ten ranking of \citet{Sezgin:2004}, one of the most referenced (therefore well-known), and implemented e.g.\ in opencv \citep{opencv_library}.

Furthermore, a quick visual check of our example images with the tool ImageJ \citep{Schneider:2012} shows, that most available other techniques lead to erroneous  binarizations, with background pixels becoming falsely detected as signals and set to white.

We analyzed one of the more promising methods further: the intermodes thresholding of \citet{Prewitt:1966} (e.g.\ implemented in ImageJ \citep{Schneider:2012}) shows a detection rate similar to that of Otsu's method. It smoothes the histogram by running averages of size 3 until there are exactly two local maxima. The threshold is then the arithmetic mean of those. But only in the example of particle separation (\autoref{subsection:Particle Separation}) the intermodes thresholding performs superior to Otsu's method, because the intermodes thresholding chooses a higher threshold than Otsu's method. The higher threshold is chosen in all our examples, and the reason for this is simple and shows also the drawback of intermodes thresholding: dominant peaks in the histogram --- such as the peak at 1 in our perfectly ``illuminated'' artificial images --- are detected as one maximum and shift the average towards the maximum brightness value. Nonetheless, the performance of such a simple approach is excellent.

In \figref{fig:example of hann filtered image}, an image is shown representing the clustering by Otsu's method. For all further steps of calculating the particle center only the threshold value detected by Otsu's method is used, not the binarized image itself.
This means that in the first step of the moment method the threshold is used to identify ``white'' pixels belonging to a particle, while in the second step the position is calculated with the brightness values of the original image.

\begin{figure}
  \includegraphics[width=0.49\textwidth]{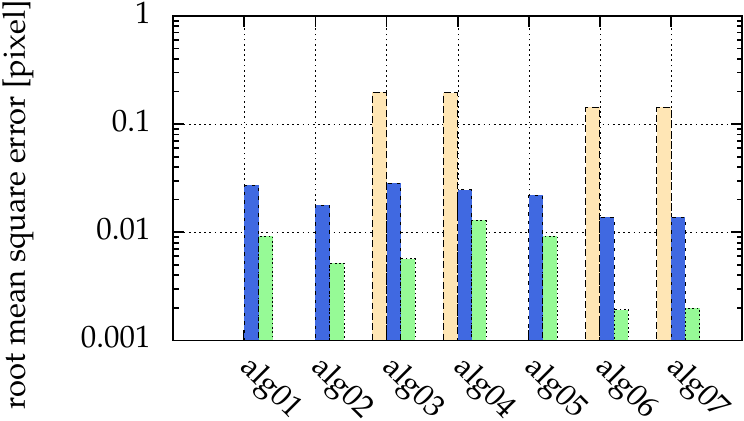}
  \caption{Comparison of different strengths of additive white Gaussian noise: \peach bars for $\SNR = 10$, \royalblue for $\SNR = 100$ and \mintgreen for $\SNR = 1000$. Missing bars imply that not all particles where correctly detected (for $SNR=5$, this was the case for all algorithms). The ordinate shows the root mean square error of the distance between detected and real positions. There was no salt and pepper noise. The statistics/simulation was done with images containing $10000$ particles with $\sigma_{\{x, y\}} = 1$.}\label{fig:noise with snr}
\end{figure}

Different algorithms are compared with respect to different signal to noise ratios in \figref{fig:noise with snr}:

\begin{description}
\item [alg01\label{alg01}] moment method (\autoref{section:Blob Detection})
\item [alg02\label{alg02}] \nameref{alg01} with $I_{base} = I_{threshold}$
\item [alg03\label{alg03}] \nameref{alg02} preprocessed by a Hanning filter ($N = 5$)
\item [alg04\label{alg04}] \nameref{alg03} with the threshold automatically detected by Otsu's method
\item [alg05\label{alg05}] \nameref{alg04} with gamma adjustment with $\gamma=3$
\item [alg06\label{alg06}] \nameref{alg01} preprocessed by a Hanning filter ($N = 5$) and fitted by a generalized Gaussian (\autoref{section:Fitting Postprocessing})
\item [alg07\label{alg07}] \nameref{alg06} with the threshold automatically detected by Otsu's method
\end{description}

\begin{figure}
  \includegraphics[width=0.49\textwidth]{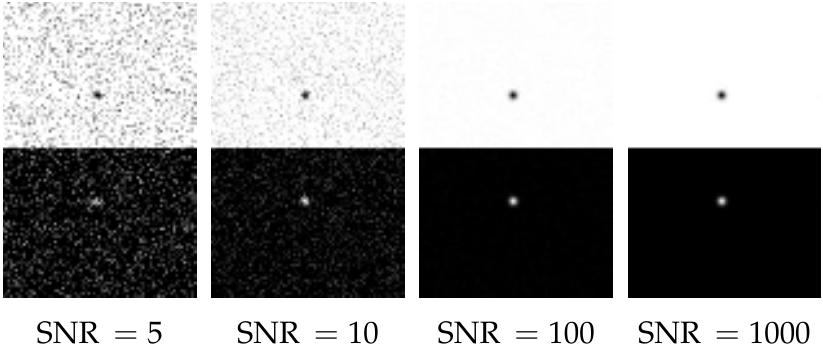}
  \caption{Example images of $2$ particles for different signal to noise ratios, which were used in \figref{fig:noise with snr}. For better visibility, the upper parts of the images are shown in inverted colors.}\label{fig:noise with snr example images}
\end{figure}

Examples of single noisy particle images are shown in \figref{fig:noise with snr example images}. For a $\SNR$ of $5$, not all of the $10000$ particles could be detected by any of the algorithms.
While particles in the example images \figref{fig:noise with snr example images} are easy to identify for human eyes, the algorithms are more sensitive to the noise.

Comparing \nameref{alg01} and \nameref{alg03} in \figref{fig:noise with snr}, we can see that the Hanning filter used in \nameref{alg03} leads to a better detection rate in the case of high noise.

\citet{Feng:2007} recommend using $I_{base} = I_{threshold}$ in the moment method to reduce uncertainties in the found particle positions. \nameref{alg02} uses this method, and \figref{fig:noise with snr} shows that indeed the error can be reduced in comparison with the pure moment method in \nameref{alg01}.
\begin{figure}
  \includegraphics[width=0.49\textwidth]{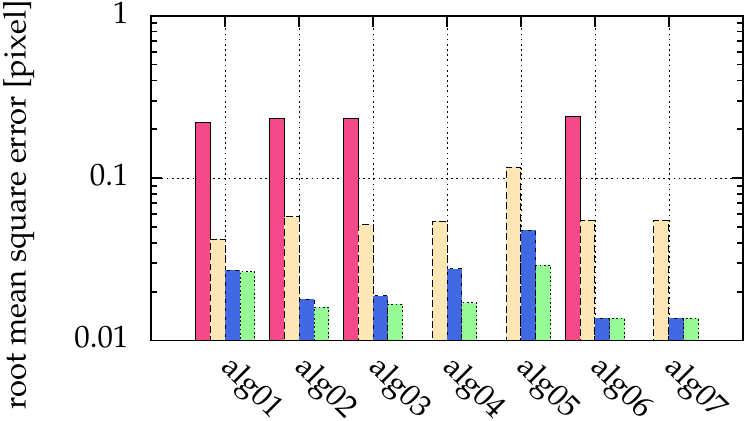}
  \caption{Comparison of different particle sizes: \frenchrose bars for $\sigma_{\{x, y\}} = 0.1$, \peach for $\sigma_{\{x, y\}} = 0.5$, \royalblue for $\sigma_{\{x, y\}} = 1$ and \mintgreen for $\sigma_{\{x, y\}} = 2$. Missing bars imply that not all particles where correctly detected. The ordinate shows the root mean square error of the distance between detected and real positions. There was a low Gaussian noise with a $\SNR = 100$, and no salt and pepper noise. The statistics/simulation was done with images containing $10000$ particles.}\label{fig:compare algorithm to particle size}
\end{figure}
\begin{figure}
  \includegraphics[width=0.49\textwidth]{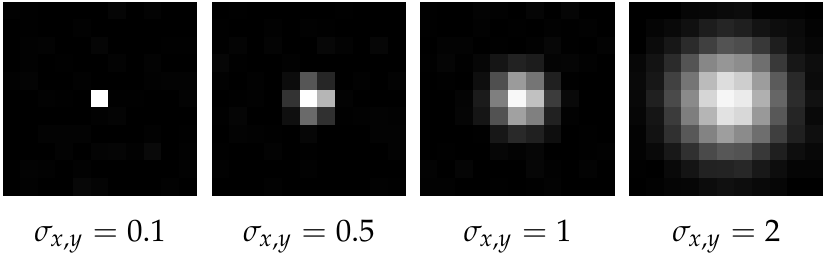}
  \caption{Example images of the different particle sizes used in \figref{fig:compare algorithm to particle size}.}\label{fig: compare algorithm to particle size example images}
\end{figure}
\begin{figure}
  \includegraphics[width=0.49\textwidth]{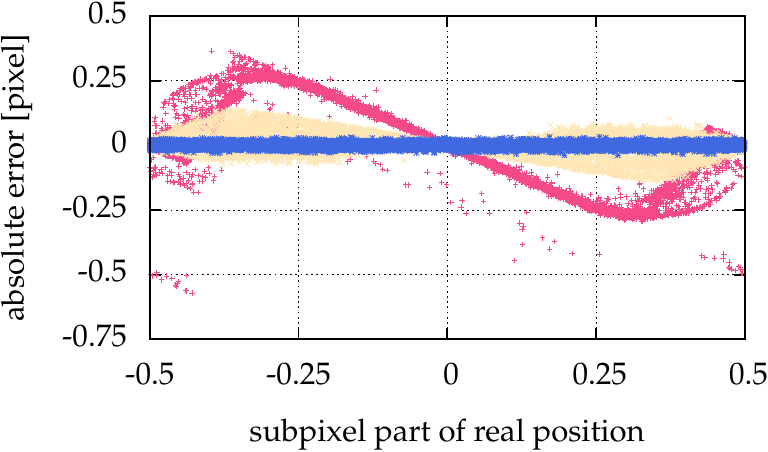}
  \caption{Comparison of different particle sizes: \frenchrose points for $\sigma_{\{x, y\}} = 0.1$, \peach for $\sigma_{\{x, y\}} = 0.5$ and \royalblue for $\sigma_{\{x, y\}} = 1$. For $\sigma_{\{x, y\}} = 2$, the same visual result was obtained as for $\sigma_{\{x, y\}} = 1$. There was a low Gaussian noise added, with a $\SNR = 100$, and no salt and pepper noise. The statistic/simulation was done with images containing $10000$ particles. The abscissa shows the sub-pixel coordinate of the real positions, whereas the ordinate shows the respective absolute error of the positions calculated with \nameref{alg06}.}\label{fig:compare algorithm to particle size positions}
\end{figure}
However, this is not true for small particles ($\sigma_{\{x, y\}} \in \{0.1, 0.5\}$), as shown in \figref{fig:compare algorithm to particle size}. While \citet{Feng:2007} explain, why an inappropriately chosen threshold leads to pixel locking, here we see that another reason for pixel locking can be missing information, such as particles consisting of not enough pixels, as seen in \figref{fig: compare algorithm to particle size example images}. This is not an error of the algorithm, but of the measurement itself.
\figref{fig:compare algorithm to particle size positions} illustrates the influence of the particle size on \nameref{alg06}.  
For particle sizes $\sigma_{\{x, y\}} \in \{0.1, 0.5\}$, the positions calculated by \nameref{alg06} are not statistically fluctuating around the real position. Instead, there is a systematic deviation depending on the real position -- a similar behavior can be observed for all presented algorithms. 
For example, a particle with $\sigma_{\{x, y\}} = 0.1$ consists of more than one pixel only, if the absolute value of the chosen sub-pixel coordinate is greater than $0.25$ (cf. \figref{fig: compare algorithm to particle size example images}).
Therefore, for a coordinate with an absolute value of the chosen sub-pixel coordinate of less than $0.25$, any algorithm can find just that one pixel, and consequently only detect the exact coordinate of it, which yields $0$ as the sub-pixel coordinate.

The clustering by Otsu's method used in \nameref{alg04} and \nameref{alg07} performs well. Only for very small particles, in the example given by \figref{fig:compare algorithm to particle size} and visualized in \figref{fig: compare algorithm to particle size example images}, a stable detection is not possible. Increasing gamma (\nameref{alg05}) does slightly improve the accuracy of \nameref{alg04}, but not all particles can be detected any more. Comparing \nameref{alg02} and \nameref{alg04} shows that Otsu's method does not choose the best threshold. But as an automatic procedure processing all available pixel values, it can reduce human errors in the process of choosing the threshold.

\section[Moment Method (Blob Detection)]{Moment Method\\(Blob Detection)}\label{section:Moment Method (Blob Detection)}

\begin{figure}
  \includegraphics[width=0.49\textwidth]{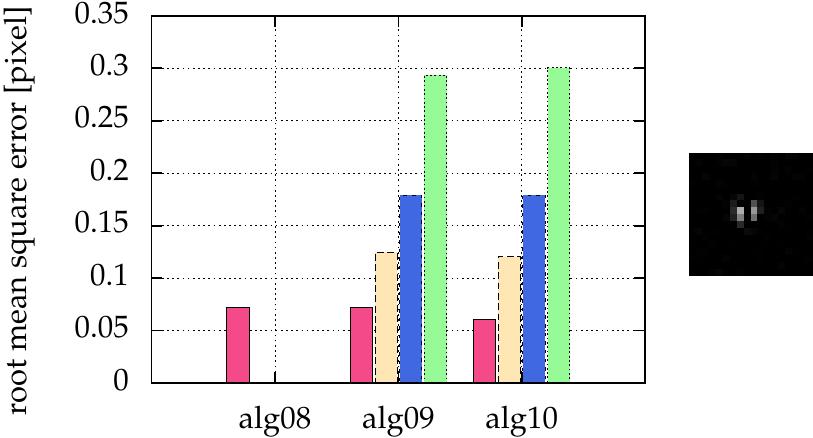}
  \caption{Comparison of different strength of pepper noise: \frenchrose bars for a probability of pepper noise of $1 \mbox{ \%}$, \peach for $5 \mbox{ \%}$, \royalblue for $10 \mbox{ \%}$ and \mintgreen for $20 \mbox{ \%}$. Missing bars imply that not all particles where correctly detected. The ordinate shows the root mean square error of the distance between detected and real positions. The Gaussian noise was added with a $\SNR = 100$. The statistics/simulation was done with images containing $10000$ particles with $\sigma_{\{x, y\}} = 1$.
The right image shows a single particle separated by pepper noise with a probability of $20 \mbox{ \%}$.}\label{fig:pepper noise}
\end{figure}

In \figref{fig:pepper noise}, different algorithms are compared with respect to different probabilities of pepper noise:

\begin{description}
\item [alg08\label{alg08}] \nameref{alg01} with a search radius $r=1$ (moment method with particles being only single connected sets, similar to \citep{Feng:2007, Ivanov:2007})
\item [alg09\label{alg09}] \nameref{alg01} with a search radius $r > 1$
\item [alg10\label{alg10}] \nameref{alg01} preprocessed by a Hanning filter ($N = 5$)
\end{description}

We can see that for high pepper noise \nameref{alg08} is not able to detect all particles correctly --- it finds too many, because some particles are split in two by the pepper noise.

Using \nameref{alg09}, the generalized moment method described in \secref{section:Blob Detection}, we are able to detect all particles correctly. The same holds for the Hanning filter in \nameref{alg10}. The quality of the latter is comparable to the generalized moment method. The only draw-back is the larger computing time of \nameref{alg10} (see \figref{fig:used processor time for noise with snr}, comparison of used processor times of \nameref{alg02} and \nameref{alg03}).

\section{Fitting (Postprocessing)}\label{section:Fitting Postprocessing}
Given approximate coordinates from the blob detection of the last \secref{section:Moment Method (Blob Detection)}, we can try to enhance them by fitting a generalized Gaussian point spread function, which is visualized in \figref{fig:example shapes} and given as:
\begin{align}
\frac{p^{1-\frac{1}{p}}}{2 \sigma \Gamma{\left( \frac{1}{p} \right)}} \times \exp{\left( \frac{-\sqrt{x^2+y^2}^p}{p \sigma^p} \right)} \label{eq:generalized Gaussian point spread function}
\end{align}

The fit is performed locally to every single particle. Therefore, we split the given image in non-overlapping squares with an approximated particle coordinate located in the center of the square. Every square is chosen with a maximal side length under the given restrictions.

Initially, the distance $d$ of two particles $i$ and $j$ is defined as:
\begin{align*}
&d\left(\left(\begin{array}{cc}x_i\\y_i\\\end{array}\right),\left(\begin{array}{cc}x_j\\y_j\\\end{array}\right)\right) := \\
&\qquad \max{\left\{|x_i-x_j|, |y_i-y_j|\right\}}
\end{align*}

Then, for a given particle coordinate $p_0 := (x_0, y_0)^T$ the closest particle $p_1 := (x_1, y_1)^T$ is found as:
\begin{align*}
&\forall i:
d\left(\left(\begin{array}{cc}x_i\\y_i\\\end{array}\right),p_0\right) \geq
d\left(p_1,p_0\right)\\
\end{align*}

With $\delta := \frac{1}{2} d((x_1, y_1)^T,(x_0, y_0)^T)$ and $\floor{.}$, $\ceil{.}$ the floor and ceiling functions\footnote{In practice, this should not be a mapping to integers $\Z$, but to image coordinates --- a subset of non-negative integers $N_0$.}
we get the vertices of the square as:
\begin{align*}
\left\{ \ceil{x_0 - \delta}, \floor{x_0 + \delta}, \ceil{y_0 - \delta}, \floor{y_1 + \delta} \right\}
\end{align*}

Now we generate separate problems for every square or particle. Let the given image be a matrix $I = \left(I_{i,j}\right)$. Here, we use the original image and not the prefiltered one.

An artificial image can be created with \eqref{eq:generalized Gaussian point spread function} as $A(x,\sigma,p) = \left(A_{i,j}(x,\sigma,p)\right)$ with the particle coordinate $x \in \R^2$ and $\sigma$, $p$ the parameters for the generalized Gaussian point spread function. With $b$ the averaged brightness of the background, this results in the optimization problem:
\begin{align*}
\min_{x,\sigma,p,b}{\sum_{i,j}^{}{\left(I_{i,j} - A_{i,j}(x,\sigma,p) - b\right)^2}}
\end{align*}

For solving this optimization problem we use the algorithm L-BFGS-B \citep{Byrd:1995, Zhu:1997, Morales:2011}, implemented in the python module/library SciPy \citep{Jones:2001}.

The gradient of the objective function is calculated numerically by a symmetric difference quotient if possible (e.g.\ on the boundary of the feasible solutions we cannot calculate a symmetric difference quotient).

\begin{figure}
  \includegraphics[width=0.49\textwidth]{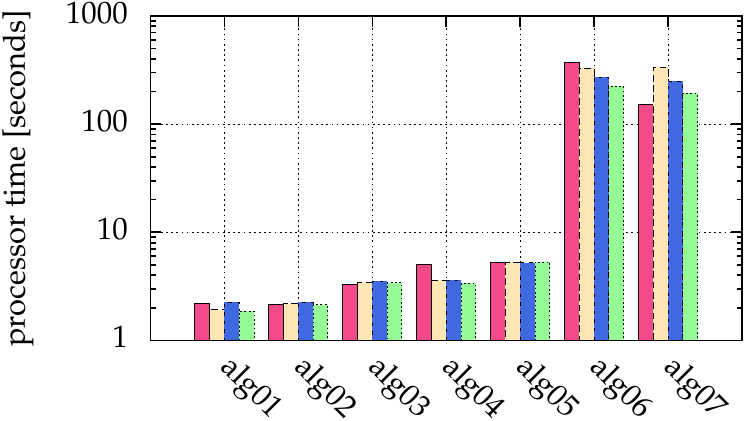}
  \caption{Comparison of used processor time of the simulations of \figref{fig:noise with snr}: \frenchrose bars for $\SNR = 5$, \peach for $\SNR = 10$, \royalblue for $\SNR = 100$ and \mintgreen for $\SNR = 1000$. The ordinate shows the measured processor time in seconds on an Intel Xeon processor E5-2643v3. The statistics/simulation was done with images containing $10000$ particles with $\sigma_{\{x, y\}} = 1$.}\label{fig:used processor time for noise with snr}
\end{figure}

In \figref{fig:noise with snr}, different algorithms were compared with respect to different signal to noise ratios, including those with fitting. Additionally, in \figref{fig:used processor time for noise with snr} the processor time used for the simulation is given. It is obvious that a small improvement by fitting a generalized Gaussian (\nameref{alg06} and \nameref{alg07}) leads to a large calculation time (\figref{fig:used processor time for noise with snr} shows a factor of $60$ to $144$).

The improved detection rate by Hanning filtering (e.g.\ \nameref{alg03}), and the automatically chosen threshold by Otsu's method (e.g.\ \nameref{alg04}) each lead to a larger error, as demonstrated in \figref{fig:noise with snr} and \figref{fig:compare algorithm to particle size}. This can be corrected by fitting (e.g.\ \nameref{alg06} and \nameref{alg07}).

Instead of successively fitting every individual particle in one image, one can try to fit all particles in one image simultaneously.
This assumption leads to a high dimensional optimization problem. In our implementation with the algorithm L-BFGS-B, this problem could not always be solved successfully. If it was successful, the result was sometimes slightly better than fitting every individual particle,  but at the cost of a considerably increased computing time: it was about $100$ times higher for $100$ particles, and about $700$ times higher for $1000$ particles.

\section{Examples}\label{section:Examples}
\subsection{Velocity}\label{subsection:Velocity}
In this section we will regard the velocity in images; this is the velocity of a particle in the image plane -- e. g. a 2D mapping of a real 3D motion.

Let us assume a sequence of images with a temporal distance of $dt = 10 \mbox{ ms}$ between $2$ consecutive images (equivalent to a frame rate of $100$ images per seconds), with particles modeled as a Gaussian with $\sigma = 1 \mbox{ pixel}$ and a $\SNR = 100$.

In our simulations  (see \figref{fig:noise with snr} and \figref{fig:compare algorithm to particle size}), the presented algorithms \nameref{alg06} and \nameref{alg07} yield a root mean square error of about $0.014 \mbox{ pixels}$ (or better). Assuming a distribution around $0$, the root mean square is the standard deviation.

When calculating a particle velocity from $2$ consecutive particle positions, each subject to the same uncertainties, error propagation leads to an error of $(2 \times 0.014)/dt$~$[\mbox{px}/\mbox{s}]$ in the velocity.

As an example, for a $4$ megapixel camera ($2048$x$2048$~$\mbox{px}^{2}$) with a field of view of $4 \mbox{ cm}$ by $4 \mbox{ cm}$ this leads to an uncertainty in the velocity of $0.056 \mbox{ mm}/\mbox{s}$.

In the following table a few examples are given with the resolution in $\mbox{mm}/\mbox{px}$, $dt$ in ms and the velocity in $\mbox{mm}/\mbox{s}$: 
\begin{center}
\begin{tabular}{|c|c|c|c|c|}\hline
\multirow{2}{*}{$\SNR$} & pixel & \multirow{2}{*}{resolution} & \multirow{2}{*}{$dt$} & velocity \\
& error &  & &  error \\\hline\hline
$100$ & $0.014$ & $0.020$ & $10$ & $0.056$\\\hline
$100$ & $0.014$ & $0.020$ & $2$ & $0.28$\\\hline
$100$ & $0.014$ & $0.005$ & $12.5$ & $0.012$\\\hline
\end{tabular}
\end{center}
We neglect here that changing the resolution also changes the size of a particle on the image sensor. Otherwise, the error for the resolution of $0.005\mbox{ mm}/\mbox{px}$ would be reduced dramatically.

With the last example representing a $4$ megapixel camera ($2048$x$2048$~$\mbox{px}^{2}$) at 80 frames per seconds with a field of view of $11 \mbox{ mm}$ by $11 \mbox{ mm}$ we would be able to measure the velocity of particles ($\varnothing = 9.19 \mbox{ $\mu$m}$, $\rho = 1.51 \mbox{ g/cm}^3$) at room temperature, which would be about $0.08 \mbox{ mm/s}$. Experiments with a crystalline 2D complex plasma, and a comparable spatial camera resolution, were analyzed with the presented \nameref{alg01} by \citet{Knapek:2007}, yielding reasonable kinetic energies of the particles.

Knowing the uncertainties, especially for particle velocity calculation, should not be underestimated: Gaussian noise can easily mask a Maxwellian velocity distribution, but it is not possible to separate the two distributions (see e.g.\ \citet[chapter 7]{Knapek:2011}; here, as well as in other publications \citep{Knapek:2007,Knapek2:2007}, the applicability of our algorithm to real-world data is demonstrated in more detail). Therefore, it is of high importance to know the limit of resolvable particle motion (depending on particle size, $\SNR$ and the algorithms) for a specific experiment before interpreting the results.

\subsection{Particle Separation}\label{subsection:Particle Separation}
\begin{figure}
  \includegraphics[width=0.49\textwidth]{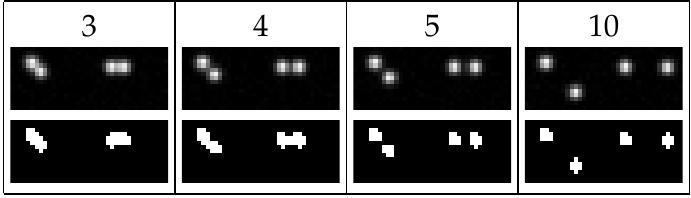}
  \caption{Different particle distances: In the first row, the minimal particle distances are given in pixel. In the second row, example images are given. In the last row, these example images are prefiltered by Otsu's method as described in \nameref{alg04}.}\label{fig:different particle distance}
\end{figure}
Here, we assume two particles which are nearby each other on the image plane (see \figref{fig:different particle distance}), e.g.\ due to their overlapping motion in different layers. Further, we assume both particles have the same size on the image, e.g.\ an uniform illumination, a good enough depth of field and same particle size and texture.

We neglect the Gaussian beam profile of the laser in the simulation. This profile might give additional information about the depth of a particle: the pixel intensity values would reflect the (ambiguous) position of the particle within the spatial extent of the laser beam, and could be used for a relative depth evaluation between particles, but we do not use this kind of information in the presented algorithms.

Now we can use all algorithms introduced in \autoref{section:Preprocessing (Image Processing)}. Since we want to separate both particles, we do not want to detect particle pictures like \figref{fig:pepper noise} (right) as one particle. Also, we cannot use a too large Hanning amplitude filter, because it would wash-out distinctive edges of nearby particles.
Therefore, we have to demand a good $\SNR$ to avoid the necessity of preprocessing. This is usually available in typical images of complex plasmas obtained with a laser filter which suppresses background illumination (e.g.\ from the plasma glow).

We now introduce additional algorithms similar to \nameref{alg03} and \nameref{alg06}:

\begin{description}
\item [alg11\label{alg11}] \nameref{alg02} with a large threshold of $190$ and a search radius $r=1$
\item [alg12\label{alg12}] \nameref{alg11} with fitting a generalized Gaussian (\autoref{section:Fitting Postprocessing})
\end{description}

\begin{figure}
  \includegraphics[width=0.49\textwidth]{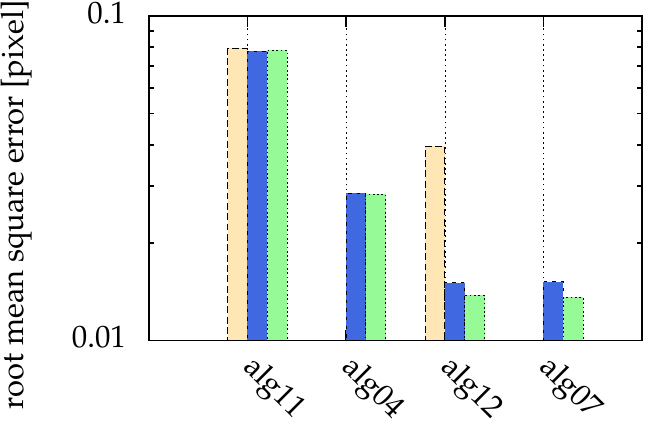}
  \caption{Comparison of different particle distances: \peach for minimal particle distance of $4$ pixel, \royalblue for minimal particle distance of $5$ pixel and \mintgreen for minimal particle distance of $10$ pixel. Missing bars imply that not all particles where correctly detected (for a distance of $3$ pixel, this was the case for all algorithms). The ordinate shows the root mean square error of the distance between detected and real positions. The statistics/simulation was done with images containing $10000$ particles with $\sigma_{\{x, y\}} = 1$ and $\SNR = 100$.}\label{fig:particle distance}
\end{figure}

In \figref{fig:particle distance}, different particle distances, as visualized in \figref{fig:different particle distance}, are compared. Here, for all presented algorithms a search radius of $r=1$ was used. Particles with distances of $3$ pixels or less could not be separated. The moment method (\nameref{alg11}) described in \autoref{section:Blob Detection} with a large threshold of $130$ is able to separate particles with $\sigma_{\{x, y\}} = 1$ with a distance of only $4$ pixels. Choosing the threshold automatically
(for minimal particle distances of $3$, $4$, $5$ and $10$ pixels, the threshold was $78$, $71$, $73$ and $73$, respectively)
with \nameref{alg04} separates these particles down to a distance of $5$ pixels. In both cases, postprocessing the images by fitting (\nameref{alg12} and \nameref{alg07}) reduced the uncertainties.

The possibility to separate close-by particles can prove helpful in 3D diagnostics, e.g.\ for the analysis of data taken with a stereoscopic setup (several cameras viewing the same volume from different angles). Particles located close to each other on the image plane are typical features for this kind of diagnostics, and algorithms are needed to reliably detect particles in each of the camera views as the basis for a subsequent triangulation\citep{Alpers:2015}.

\section{Conclusion}\label{section:Conclusion}
In this paper we presented a comparison of several methods and algorithms for particle tracking from images. The methods and algorithms were tested on artificial images simulating data as they are obtained in complex plasma experiments, including realistic image noise (additive white Gaussian noise, salt and pepper noise). To increase the statistical significance, images with a large number of particles ($10000$) were analyzed. The proposed procedure for particle tracking consists of three major steps: image processing, blob detection and postprocessing.

In \autoref{section:Preprocessing (Image Processing)}, we show that using a Hanning filter to remove Gaussian noise during image processing results in a better detection rate in the presence of high noise, whereas the accuracy of the found positions is slightly reduced (\figref{fig:noise with snr}).

For images consisting of features (the particles) and a background (noise), the choice of a good threshold is important during image processing. With Otsu's method (used in \nameref{alg04}, \nameref{alg07}), we introduce this concept of automatic thresholding for particle detection in complex plasma for the first time (\autoref{section:Preprocessing (Image Processing)}). Other automatic thresholding techniques were tested, but did not prove to be suitable. The clustering by Otsu's method performs very well (\figref{fig:noise with snr}), yielding almost the same results as the manually chosen threshold for all but the smallest particle sizes (\figref{fig:compare algorithm to particle size}).
On the one hand, choosing the right threshold value is not an easy task, and an automatic method can dramatically reduce human errors. On the other hand, an automatism prohibits using expert knowledge of the user in special circumstances, e.g.\ for the task of particle separation (\autoref{subsection:Particle Separation}).

In \autoref{section:Blob Detection} we introduce an improved algorithm for blob detection: we generalized the set used for the moment method to a not necessarily simply connected set, and show that we can considerably improve particle detection in the presence of certain kinds of noise (e.g.\ salt and pepper noise, \figref{fig:pepper noise}) with this generalization.

We present a postprocessing method in \autoref{section:Fitting Postprocessing} to further enhance the accuracy of the detected particle positions by fitting a generalized Gaussian function to the intensity profiles of the particles. This is in particular interesting if prefiltering is necessary due to noisy images. Then, the postprocessing can reduce errors introduced by the prefilter (\figref{fig:compare algorithm to particle size}, \figref{fig:noise with snr}). Also, it can increase the sub-pixel resolution of particle positions. This is especially interesting for applications where small particle velocities, e.g.\ thermal velocities, are calculated from the positions (\autoref{subsection:Velocity}).

Another application is shown in \autoref{subsection:Particle Separation}: Particles which are close-by each other on the image plane can be separated by either manual or automatic threshold detection, and position accuracy was improved by the above postprocessing method. This kind of situation typically appears in the individual camera images of a stereoscopic imaging system. 

In summary, image processing with a Hanning filter (\nameref{alg03}), and a subsequent blob detection with the moment method detects in the most cases all particles in our simulations, but needs a manually chosen threshold. Automatic threshold detection (\nameref{alg04}) results in a slightly reduced accuracy and a reduced detection rate, but has the advantage of the automatism. In both cases, postprocessing the acquired positions by fitting (\nameref{alg06} and \nameref{alg07}) reduced uncertainties in the particle coordinates at the cost of a large calculation time (\figref{fig:used processor time for noise with snr} shows a factor of $60$ to $144$), but for specific experiments with the requirement of a good sub-pixel resolution this can be very useful and worth the effort. 

\section{Acknowledgments}
The authors are funded by DLR/BMWi under grant FKZ 50WM1441, and by STMWi (Bayerisches Staatsministerium f{\"u}r Wirtschaft und Medien, Energie und Technologie).

\bibliographystyle{plainnat}
\bibliography{literature}

\end{document}